\documentclass[12pt]{article}
\usepackage{color,epsfig,amsmath,amssymb,graphicx}

\font\srm=cmr8

\definecolor{violet}{rgb}{0.4,0,0.6}
\definecolor{vert}{rgb}{0,0.6,0.2}
\definecolor{navy}{rgb}{0.0,0.0,0.4}

\definecolor{bleu}{rgb}{0.3,0.0,0.8}
\definecolor{brun}{rgb}{0.6,0.3,0.0}

\def\spose#1{\hbox to 0pt{#1\hss}}\def\lta{\mathrel{\spose{\lower 3pt\hbox
{$\mathchar"218$}}\raise 2.0pt\hbox{$\mathchar"13C$}}}  \def\gta{\mathrel
{\spose{\lower 3pt\hbox{$\mathchar"218$}}\raise 2.0pt\hbox{$\mathchar"13E$}}}

\def\Go{ {g}}            \def\Xo{ {x}}
\def\gg{ {\color{blue} \overline g}}
\def\per{ {\color{blue} \perp}}
\def\gamm{ {\color{blue} \gamma}}
\def\lamda{ {\color{blue} \lambda}}
\def\nnu{ {\color{blue} \nu}}
\def\vv{ {\color{blue} v}}
\def\ww{ {\color{blue} w}}
\def\nab{ {\color{blue} \overline \nabla}}         
\def\xx{ {\color{blue} \overline x}}         
\def\xx{ {\color{blue} \overline x}}
 
\def\calI{ {\color{brun} {\cal I}}}         
\def\LL{ {\color{brun} L}}
\def\rrho{ {\color{brun} \rho}}
\def\TT{ {\color{brun} T}}
\def\PP{ {\color{brun} P}}
\def\Bbeta{ {\color{brun} \beta}}
\def\EE{ {\color{brun} E}}
\def\CC{ {\color{brun} {\mathfrak C}}}
\def\calK{ {\color{brun} {\mathfrak K}}}
\def\OM{ {\color{brun} \Omega}}
\def\OP{ {\color{brun} {\mathfrak O}}}
\def\Hh{ {\color{brun} {\mathfrak H}}}
\def\Hhh{ {\color{blue} \overline{\color{brun}\mathfrak H}}}

\def\xe{{\color{vert}\xi}} \def\ze{{\color{vert}\zeta}}
\def\ete{{\color{vert}\eta}}

\def\Ke{{\color{blue}K}}
\def\delte{{\color{blue}\delta}}

\def\Qq{ {\color{violet} {\cal Q}} }
\def\vphi{ {\color{violet} \varphi} }
\def\muv{ {\color{violet} \mu} }

\def\uu{ {\color{violet} u} }

\def\be{\begin{equation} }
\def\fe{\end{equation}}
\def\ee{\end{equation}}
\newcommand{\ba}{\begin{eqnarray}}
\newcommand{\ea}{\end{eqnarray}}

\begin{document}
\bigskip

\centerline{Contribution to {\it Micro and Macro Structures of Spacetime},} 
\centerline{ 10th Peyresq Physics meeting, June 2005.}
\bigskip
\bigskip
\bigskip

\begin{center}
{\Large \textcolor{red}{\Large Poly-essential and general
Hyperelastic\\ World (brane) models.}}
\end{center}
\bigskip
\bigskip
\bigskip
\centerline{\large
\textcolor[named]{ForestGreen}{Brandon Carter}$^\sharp$}
\bigskip
\bigskip
\centerline{ {$^\sharp$}LuTh, Observatoire de Paris-Meudon, 92195
Meudon, France.} 
\bigskip
\bigskip
\centerline{November 2005}

\vskip 3 cm

\begin{abstract}
This article provides a unified treatment of an extensive
category of non-linear classical field models whereby the 
universe is represented (perhaps as a brane in a higher 
dimensional background) in terms of a structure of a 
mathematically convenient type describable as hyperelastic, 
for which a complete set of equations of motion is provided 
just by the energy-momentum conservation law.  Particular 
cases include those of a perfect fluid in quintessential 
backgrounds of various kinds, as well as models of the 
elastic solid kind that has been proposed to account for 
cosmic acceleration. It is shown how an appropriately 
generalised Hadamard operator can be used to construct a 
symplectic structure that controles the evolution of small
perturbations, and that provides a characteristic equation
governing the propagation of weak discontinuities
of diverse (extrinsic and extrinsic) kinds. The special
case of a poly-essential model - the k-essential analogue
of an ordinary polytropic fluid - is examined and shown
to be well behaved (like the fluid) only if the pressure
to density ratio $w$ is positive.

\end{abstract}

\vfill\eject

\bigskip\noindent
 {\bf 1. Introduction.}
\medskip

As a generalisation of the category of media that are elastic in
the ordinary variational sense \cite{C73,FS75}, the extensive 
category of models referred to here as ``hyperelastic'' is 
characterised by an action density $\LL$ that can be formulated as a 
non-linear function just of a set of scalar $p+1$ scalar fields 
$\vphi^{_{\,0}}\, ,\vphi^{_1}\, , ... \vphi^{p} $ and their gradient 
components $\vphi^{_0}_{\, ,a}\, ,\vphi^{_1}_{\, ,a}\, , ... 
\vphi^{p}_{\, ,a} $ with respect to coordinates $\xx^a$ $\,(a=0,1, ... 
p)\,$ on a $p+1$ dimensional worldsheet with codimension $q\geq0$ in a
background space time endowed with  Lorentz signature metric that has
components $\Go_{\mu\nu}$ with respect to coordinates $x^\mu$,
$(\mu =0,1, ..., p+q)$.   

This work is intended for the treatment of scenarios of the usual 
cosmological type with space dimension $p=3$, and it is set up 
(using a background tensor formalism of the kind \cite{C00} that
was originally developed for the treatment of conducting cosmic 
strings \cite{CS04}) in such a way as to be applicable not just to 
models of the traditional kind in which the background spacetime 
dimension is 4, so that the codimension $q$ vanishes, but also to 
models of more exotic ``brane world'' varieties, in which the 
background is of higher dimension, 5 or more.

In a typical cosmological  application of such a model, $\vphi^{_{\,0}}$ 
would represent a ``quintessence scalar'' of the kind commonly invoked
\cite{DCS03}, to account for the apparent observation of cosmic 
acceleration while the other scalars $\vphi^{_1}\, ,$ 
$\vphi^{_2}\, ,$ $\vphi^{_3}\,$ would be interpretable as comoving 
(Lagrange type) coordinates of a material medium of the ``normal''
kind which in the simplest case would be of ordinary perfect fluid 
type. However instead of being a perfect fluid, the ``normal'' matter
characterised by such comoving coordinates could just as well be an 
elastic solid of the kind envisaged by Bucher and Spergel 
\cite{BS99,C04,BCCM05}.

In a non-cosmological application for which a model of this 
hyperelastic kind might be used, the ``normal'' constituent could be 
that of a solid neutron star crust, within which a freely flowing
 superfluid neutron current would be characterised by the scalar
field, whose gradient would be the neutron momentum covector, 
$\muv_a=\vphi^{_0}_{\, ,a}$. In that case the scalar would have to be 
of ignorable type (meaning that the Lagrangian would depend just on 
its gradient but not on its undifferentiated value $\vphi^{_0}$) so 
that, according to equation  (\ref{concur}) below, the neutron
current would automatically be conserved. (It is however to be remarked 
that in a realistic treatment of a neutron star crust it may be 
necessary to allow for the possibility \cite{LSC98} that the superfluid 
neutron constituent may not be separately conserved, so that a more 
elaborate kind of model would then be needed.)

Whatever its dimension, the background spacetime metric will induce a 
corresponding worldsheet metric with components
{\be \gg_{ab}={\Go}_{\mu\nu}\Xo^{\mu}_{\, ,a}\Xo^{\nu}_{\, ,b}
\, ,\fe}
and with determinant $\vert\gg\vert$ in terms of which the
action integral will be expressible as
{\be \calI=\int \LL\,\Vert\gg\Vert^{1/2}\, {\rm d}^{p+\!1}\xx\, .\fe}

In order for the system to be considered as regular, the induced metric 
on the worldsheet must of course itself have a Lorentz signature, and 
furthermore the scalar field gradients must be linearly independent so
that the fields themselves will be adoptable as an admissible set of 
worldsheet coordinates, for which we shall simply have $\xx^a=\vphi^a$, 
which entails that the Lagrangian density $\LL$ will depend just on 
the undifferentiated position coordinates $\vphi^a$ and on the 
corresponding set of $p(p+\!1)/2$ induced metric components $\gg_{ab}$. 

A subcategory of particular interest consists of models for which the 
Lagrangian contributions from ``scalar'' and ``normal'' parts 
separate as a sum of the form $\LL=\LL_{\rm_{S}}+\LL_{\rm_{N}}\, ,$ 
in which the ``normal'' part $\LL_{\rm_{N}}$ is independent of 
$\vphi^{_{\,0}}$ and of $\gg_{_0a}\, ,$ while the ``scalar'' part
is a (non-linear) function only of $\vphi^{_{\,0}}$ and of its squared 
gradient as given by the single induced metric component $\gg_{_{00}}$. 

\bigskip
{\bf 2. Dynamics of a regular hyperelastic system.}
\medskip

Whichever coordinate system is used, the worldsheet stress energy density
tensor will have components given \cite{C00} by
{\be \TT^{ab}=2\Vert\gg\Vert^{-1/2}\frac{\partial\big(\Vert\gg\Vert^{1/2} 
\LL\big)}{ \partial \gg_{ab}}\, .\label{stressen}\fe}
and there will be a corresponding  world hyper-elasticity tensor 
defined on the worldsheet in the manner introduced by Friedman and 
Schutz \cite{FS75} as
{\be{\CC}{^{abcd}}=\Vert\gg\Vert^{-1/2}\frac{\partial\big(\Vert\gg
\Vert^{1/2}\TT^{ab} \big)}{\partial g_{cd}} ={\CC}{^{cdab}}
\, ,\label{hyperel}\fe}
which (modulo a proportionality factor of -2) is interpretable 
as a relativistic extension of an  ordinary (purely spacelike) 
Cauchy type elasticity $E^{abcd}$ such as will be discussed below.

These worldsheet tensors will map naturally into corresponding 
background space time tensors given by the expressions
{\be \TT^{\mu\nu}=\TT^{ab}\Xo^{\mu}_{\, ,a}\Xo^{\nu}_{\, ,b}=
2\,\frac {\partial\LL}{ \partial \gg_{\mu\nu}} +\LL \,\gg^{\mu\nu} 
 \, ,\fe}
{\be \CC^{\mu\nu\rho\sigma}=\CC^{abcd}\Xo^{\mu}_{\, ,a}
\Xo^{\nu}_{\, ,b}\Xo^{\rho}_{\, ,c}\Xo^{\sigma}_{\, ,d}=\,\frac
{\partial\TT^{\mu\nu}}{ \partial \gg_{\rho\sigma}} +\frac{_1}{^2}
\TT^{\mu\nu}\gg^{\rho\sigma}  \, ,\label{hyperC}\fe}
in which $\gg^{\mu\nu}$ denotes the (first) fundamental tensor
of the worldsheet as defined \cite{C00} in terms of the contravariant
version $\gg^{ab}$ of the induced metric by the formula
{\be \gg^{\mu\nu}=\gg^{ab}\Xo^{\mu}_{\, ,a}\Xo^{\nu}_{\, ,b}
\label{fun}\, .\fe}
When the codimension $q$ vanishes the overline is redundant here as
(\ref{fun}) will then just give back the contravariant version
$\Go^{\mu\nu}$ of the background spacetime metric, and the corresponding
mixed version $\gg{^{\,\mu}_{\ \nu}}$ will then be just the same as the 
Kronecker unit tensor $\delte^\mu_\nu$, but in a background of higher
dimension $n>p+\!1$ this mixed version $\gg^\mu_{\, \nu}$
will be a non trivial (rank $p+\!1$)  projector mapping vectors onto
their world sheet tangential parts, and giving a corresponding
world sheet gradient operator
{\be \nab_{\!\nu}=\gg{^{\,\mu}_{\ \nu}}\nabla_{\!\mu}\, ,\fe}
in which the distinguishing overline is again redundent in, but only in,  
the case of vanishing codimension $q$.

As in more general brane models \cite{C00}, when the variational field 
equations ensuring invariance of the action with respect to localised 
perturbations of the worldsheet and the dynamical fields theron are 
satisfied, it will automatically follow as a Noether identity that the 
stress energy given by (\ref{stressen})  will satisfy a divergence 
condition of the standard form
{\be \nab_{\!\mu}\TT^\mu_{\ \nu}=0\, .\label{ccon}\fe}

What distinguishes hyperelastic models from others of a more general 
kind is that in the hyperelastic case no other evolution equations are 
needed: by itself (\ref{ccon}) is not only necessary but will also be 
sufficent to ensure that the variational field equations are all 
satisfied.

The way this works is that, to start with, the evolution of the world 
sheet location (which will only be needed if the codimension $q$ is 
non-zero) will be governed as always \cite{C00} by the orthogonal 
projection of (\ref{ccon}) which will take the standard form
{\be \TT^{\mu\nu}\Ke_{\mu\nu}{^{\!\rho}}=0\, ,\label{orthof}\fe}
in which $\Ke_{\mu\nu}{^{\!\rho}}$ is the second fundamental
tensor as defined~\cite{C00} by
{\be \Ke_{\mu\nu}{^\rho}= \gg{^{\,\sigma}_{\ \nu}}
\nab_{\!\mu}\gg{^{\,\rho}_{\ \sigma}}\, .\fe}
As well as the Weingarten integrability condition
 $\Ke_{\mu\nu}{^\rho}=\Ke_{\nu\mu}{^\rho}$ this defining relation
ensures the worldsheet orthogonality condition
$\Ke_{\mu\nu}{^\rho}\,\gg{^{\,\sigma}_{\ \rho}}=0$, which 
evidently entails that $\Ke_{\mu\nu}{^\rho}$ itself will
vanish in the traditional $q=0$ case, for which there are no
external dimensions so that $\gg{^{\,\sigma}_{\ \rho}}$ will just be
the identity matrix $\delte^\sigma_{\,\rho}$.

When the condition (\ref{orthof}) (involving $q$ independent 
component equations) has been satisfied, the remaining part
of (\ref{ccon}) will consist just of its tangentially projected
part, namely the set of $p+\!1$ internal component equations
given by  
{\be \gg{^{\,\nu}_{\ \rho}} \nab_{\!\mu}\TT^\mu_{\ \nu}=0\, ,\fe}
which is just what is needed to determine the evolution of
the $p+\!1$ independent worldsheet scalars $\vphi^a$.

\bigskip
{\bf 3. Symplectic perturbation currents and characteristic equation.}
\medskip

The sufficiency of the divergence condition (\ref{ccon})
as a complete set of equations of motion can be understood
as a consequence of the feature that with respect to a
 reference system of the prefered kind in which
the scalars $\phi^a$ are used directly as worldsheet
coordinates, the configuration of the system will be fully
determined just by the specification of the background
coordinates $\Xo^\nu$ as fonctions of these scalars.

It follows that, with respect to such a preferred system,  a 
perturbation of the configuration will be fully determined
just by the specification of the corresponding background
coordinate displacement, $\delta\Xo^\mu=\xe^\mu$ say, which, to 
be dynamically admissible, must of course be such as to satisfy 
the linear evolution equation obtained as the first order 
perturbation of (\ref{ccon}). When another particular solution, 
$\delta\Xo^\mu=\ete^\mu$ say, is already available (for example 
as a trivial perturbation generated by a symmetry of the system) 
then the linear equation governing a generic perturbation
$\xe^\mu$ will be conveniently expressible as the conservation, 
{\be \nab_{\!\nu}\OM^\nu=0\, ,\label{symcon}\fe}
of a symplectic worldsheet surface current (such as is also of 
interest \cite{CF02} for the purpose of quantisation) of the 
kind that has recently been shown to be straightforwardly 
constructable for a widely extended category of brane systems
\cite{C03}.  As in the case of applications \cite{CS04} to 
conducting cosmic strings, so also in the more general 
hyperelastic systems considered here, the possibility of 
expressing the perturbation just in terms of a displacement 
vector $\xe^\mu$ allows the symplectic current $\OM^\nu$ to be 
given explicitly by an expression of the form
{\be \OM^\nu\{\vec\xe,\vec\ete\}=\ete^\mu\OP_{\!\mu}^{\ \nu}
\{\vec\xe\}-\xi^\mu\OP_{\!\mu}^{\ \nu}\{\vec\ete\}\, ,\label{symp}\fe}
in terms of what I shall refer to as the hyper-Hadamard operator,
which is a linear differential operator whose action on the vector 
field $\vec \xe $ is given by a prescription of the form
{\be \OP_{\!\mu}^{\ \nu}\{\vec\xe\}=\Hh{_{\mu\ \rho}^{\ \,\nu\ \sigma}}
\nab_{\!\sigma}\xe^\rho\, .\fe}
in terms of a corresponding hyper-Hadamard tensor that can be seen
\cite{C03} to be given in terms of the hyper Cauchy tensor
(\ref{hyperC}) by the formula
{\be \Hh{_{\mu\ \rho}^{\ \,\nu\ \sigma}}=\Go_{\mu\rho}
{\TT}{^{\nu\sigma}}+2{\CC}{_{\mu\ \rho}^{\ \,\nu\ \sigma}}\, .\fe}

It is to be remarked that the standard decomposition
{\be \Go_{\mu\rho}=\gg_{\mu\nu}+\per_{\mu\nu}\, ,\fe}
 of the background metric into respectively worldsheet tangential and
 worldsheet orthogonal parts $\gg_{\mu\nu}$ and $\per_{\mu\nu}$ (of 
which the latter will vanish when the codimension $q$ is zero) 
will engender a corresponding decomposition
{\be \Hh{_{\mu\ \rho}^{\ \,\nu\ \sigma}}=
\Hhh{_{\mu\ \rho}^{\ \,\nu\ \sigma}}+
\Hh{_{\mu\ \rho}^{\per \,\nu\ \sigma}}\, ,\label{hyde}\fe}
in which the world sheet tangential part is given by
{\be \Hhh^{\mu\nu\rho\sigma}=\Hhh^{abcd}\Xo^{\mu}_{\, ,a}
\Xo^{\nu}_{\, ,b}\Xo^{\rho}_{\, ,c}\Xo^{\sigma}_{\, ,d}\, ,\hskip 1 cm
\Hhh^{abcd}=\gg^{ac}\TT^{bd}+2\CC^{abcd}\, ,\label{Cauch}\fe}
and the remainder in (\ref{hyde}) is given simply by
{\be\Hh{_{\mu\ \rho}^{\per \,\nu\ \sigma}}=\per_{\mu\rho}\,
\TT^{\nu\sigma}\, .\label{perpH}\fe}

As in the simple elastic case \cite{C72}, one can obtain the 
characteristic equation governing the propagation of a discontinuity
across a worldsheet hypersurface with normal covector 
{\be \lamda_a=\lamda_\mu\,\Xo^{\mu}_{\, ,a}\, ,\hskip 1 cm
\lamda_\mu\per^{\!\mu}_{\,\nu}=0\, ,\fe}
of the second derivative of the perturbation vector $\vec\xe$,
using the Hadamard rule to the effect that it must be given by
an expression the form
{\be \left[\nab_{\!\mu}\nab_{\!\nu}\xe{^\rho}\right]=\lamda_\mu
\lamda_\nu\ze{^\rho}\, ,\fe}
in which the vector $\vec\ze$ is a measure (whose calibration
depends on the normalisation of the -- possibly null -- characteristic
covector $\lambda_a$) of the amplitude of the discontinuity.
It can be seen to follow from this rule that the discontinuity
of the divergence of the symplectic current (\ref{symp}) will have
the form 
{\be \left[\nab_{\!\nu}\OM^\nu\{\vec\xe,\vec\ete\}\right]=\ete^\mu
\Hh{_{\mu\ \rho}^{\ \,\nu\ \sigma}}\lamda_\nu\lamda_\sigma
\ze^\rho\, .\fe}
Since the the conservation law (\ref{symcon}) is applicable to
an arbitrary reference perturbation solution $\vec\ete$ (for which
undifferentiated components $\ete^\mu$ may be chosen without 
restriction at any single given point) it can be seen that the
characteristic covector $\lamda_\mu$ must satisfy the condition
{\be \Hh{_{\mu\ \rho}^{\ \,\nu\ \sigma}}\lamda_\nu\lamda_\sigma
\ze^\rho=0\, .\fe}

For extrinsic perturbations of the worldsheet location, as obtained
by taking $\vec\ze$ to be worldsheet orthogonal, it can be seen
from (\ref{perpH}) that we simply recover the result (which is 
already known \cite{C00} to apply for a generic, not just hyperelastic,
brane model) that the characteristic covector must be a null 
eigenvector of the stress energy tensor:
{\be \gg_{\mu\nu}\ze^\nu=0\hskip 1 cm \Rightarrow \hskip 1 cm
\TT^{\mu\nu}\lamda_\mu\lamda_\nu=0\, .\label{extrin}\fe}
On the other hand for tangential perturbations -- the only kind
that can exist when the codimension $q$ is zero -- the 
characteristic equation will be expressible in purely
worldsheet tensorial form:
{\be \ze^\mu=\Xo^{\mu}_{\, ,a}\ze^a\hskip 1 cm \Rightarrow 
\hskip 1 cm  \Qq_{ac}\ze^c=0\, ,\fe}
with
{\be \Qq_{ac}=\Qq_{ca}=\Hhh_{a \ c}^{\ b\ d}\lamda_b\lamda_d\, .
\label{charm}\fe}
The condition for $\lamda_a$ to be an intrinsic characteristic
covector is thus that the determinant $|\Qq|$ of this symmetric
matrix (\ref{charm}) should vanish.

\bigskip
{\bf 4. Standard flow decomposition.}
\medskip

To qualify as hyperelastic in the strictest sense, the system should 
include a subsystem of ordinary elastic type, as characterised by the 
requirement that all the scalar fields except one, $\vphi^{_{\,0}}$ say, 
should have spacelike gradients and therefore timelike worldsheets that 
intersect on a congruence of timelike worldlines, whose unit world 
tangent vector, with components $\uu^a$ say, will be specified, for 
$\hbox{\srm A}=1, ..., p\, ,$ by
{\be \uu^a\vphi^{_A}_{\, ,a}=0\, ,\hskip 1 cm 
\gg_{ab}\uu^a\uu^b=-1\, .\fe}
Subject to the understanding that the small letters refer to an
arbitrary worldsheet coordinate system, but that the capitals refer
to a system of the preferred type specified by setting
$\xx^{_0}=\vphi^{_{\,0}}$ and $\xx^{_A}=\vphi^{_A}$, the induced metric 
components on which the action depends can then be listed in a system of 
the latter type as
{\be \gg^{_{00}}=\gg^{ab}\vphi^{_{\,0}}_{\ ,a}\vphi^{_{\,0}}_{\ ,b}
\, , \hskip 1 cm \gg^{_{A0}}=\gg^{ab}\vphi^{_A}_{\, ,a}
\vphi^{_{\,0}}_{\ ,b}\, ,\fe}
and
{\be \gg^{_{AB}}=\gg^{ab}\vphi^{_A}_{\, ,a}\vphi^{_B}_{\, ,a}
=\gamm^{ab}\vphi^{_A}_{\, ,a}\vphi^{_B}_{\, ,a}=\gamm^{_{AB}}
\label{space}\, ,\fe}
in which the space projected part of the metric is specified in
of the usual way as
{\be \gamm^{ab}=\gg^{ab}+\uu^a\uu^b\, .\fe}
The generic action variation will be expressible in terms of these
quantities in the form
{\be \delta\LL=\frac{\partial\LL}{\partial\vphi^{_{\,0}}}
\,\delta \vphi^{_{\,0}}+\frac{\partial\LL}{\partial\vphi^{_A}}
\,\delta \vphi^{_A}+\LL_{_{00}}\,\delta\gg^{_{00}}+2\LL_{_{A0}}
\,\delta\gg^{_{A0}}+\LL_{_{AB}}\,\delta\gg^{_{AB}} \, ,
\label{xpan}\fe}
which provides the components with respect to the preferred system
of the worldsheet tensor 
{\be\LL_{ab}=\frac{\partial\LL}{\partial \gg^{ab}}\fe}
whose contravariant version $\LL^{ab}=\gg^{ac}\gg^{bd}\LL_{cd}$
provides the partial derivatives
{\be \frac{\partial\LL}{\partial \gg_{ab}}=-\LL^{ab}\fe}
that are  needed for the evaluation of the expression (\ref{stressen}) 
for the stress energy tensor, which works out as
{\be \TT^{ab}=\LL\gg^{ab}-2\LL^{ab}\, .\label{stren}\fe}
At the next order one obtains the tensor
{\be\LL_{abcd}=\frac{\partial\LL_{ab}}{\partial \gg^{cd}}=\frac{
\partial^2\LL}{\partial \gg^{ab}\partial\gg^{cd}}\fe}
which provides the partial derivatives
{\be \frac{\partial^2\LL}{\partial \gg_{ab}\partial\gg_{cd}}
=\LL^{abcd}+\LL^{a(c}\gg^{d)b}+\gg^{a(c}\LL^{d)b}\fe}
that are needed for the evaluation of the expression (\ref{hyperel})
for the hyper-Cauchy tensor, which is thereby obtained as
{\be\CC^{abcd}=2 (\LL^{abcd}\!+\!\LL^{a(c}\gg^{d)b}\!+\!
\gg^{a(c}\LL^{d)b})\!-\!(\LL^{ab}\gg^{cd}\!+\!\gg^{ab}
\LL^{cd})\!+\!\frac{\LL}{2}(\gg^{ab}\gg^{cd}\!-\!2\gg^{a(c}\gg^{d)b})
\, .\label{hyp}\fe}
According to (\ref{Cauch}) the hyper-Hadamard tensor will therefore 
be given (using square brackets for index antisymmetrisation) by the 
expression
{\be \Hhh^{abcd}=4\LL^{abcd}+2\LL^{ac}\gg^{db}+2\LL^{a[d}\gg^{b]c}
+2\gg^{a[d}\LL^{b]c}+\LL\gg^{a[d}\gg^{b]c}\, ,\fe}
from which it can be seen that the characteristic matrix (\ref{charm}) 
will be obtained as
{\be \Qq_{ac}=2(\LL_{ac}\,\gg^{bd}+2\LL_{a\ c}^{\,\ b\ d})\lamda_b\lamda_d
\, .\label{charmat}\fe}

The expansion (\ref{xpan}) provides a corresponding Eulerian 
variation -- as carried out for an undisplaced worldsheet location
at a fixed value of the background coordinates $\Xo^\mu$ and metric 
$\Go_{\mu\nu}$  -- that will be expressible in the form
{\be \delta_{\rm_E}\LL=\frac{\partial\LL}{\partial\vphi^{_{\,0}}}
\,\delta \vphi^{_{\,0}}\!+\frac{\partial\LL}{\partial\vphi^{_A}}
\,\delta \vphi^{_A}\!+ J_{_0}^{\, a}
\delta\vphi^{_{\,0}}_{\, ,a}\!+J_{_A}^{\, a}
\delta\vphi^{_{A}}_{\, ,a} \, ,\fe}
in terms of current vectors given by 
{\be J_{_0}^{\, a}=2\gg^{ab}\ (\LL_{_{00}}\vphi^{_{\,0}}_{\, ,b}
\!+\!\LL_{_{A0}}\vphi^{_{A}}_{\, ,b})\, , \hskip 1 cm J_{_A}^{\, a}=
2\gg^{ab}
(\LL_{_{A0}}\vphi^{_{\,0}}_{\, ,b}\!+\!\LL_{_{AB}}\vphi^{_{B}}_{\, ,b})
\,  ,\label{currents}\fe}
which, according to the variational field equations, will have to
satisfy worldsheet divergence conditions of the form 
{\be \nab_{\!a}J_{_0}^{\, a}=\frac{\partial\LL}{\partial
\vphi^{_{\,0}}}\, ,\hskip 1 cm  \nab_{\!a}J_{_A}^{\, a}=
\frac{\partial\LL}{\partial\vphi^{_A}}\, .\label{concur}\fe}

\bigskip
{\bf 5. The separated case.}
\medskip

In many of the applications of interest a substantial simplification
will be provided by the separation of the action density as a sum of 
the form
{\be \LL=\LL_{\rm_{S}}+\LL_{\rm_{N}}\fe}
in which the ``normal'' part, $\LL_{\rm_{N}}$, is independent of 
$\vphi^{_{\,0}}$, and $\vphi^{_{\,0}}_{\, ,a}$ so that it has the form 
of the action density of an ordinary elastic medium as specified 
\cite{C73,C83} as a function just of $\vphi^{_A}$ and $\gg^{_{AB}}$, 
while the remainder $\LL_{\rm_{S}}$ is just the action of a simple, 
albeit non-linear, scalar field model, means that it depends only on 
$\vphi^{_{\,0}}$ and on the magnitude, $\muv$ say, of its gradient 
1-form
{\be \muv_a=\vphi^{_{\,0}}_{\ ,a}\, ,\fe}
as given by  
{\be \gg^{_{00}} = \gg^{ab}\muv_a\muv_b=-\mu^2\, .\fe} 

In such a system, neither part will have any dependence on 
$\gg^{_{A0}}$, so the cross terms (providing the effect known as 
``entrainment'') in the currents (\ref{currents}) will vanish, 
{\be \LL_{_{A0}}=0\, .\fe}
The two subsystems will thus be effectively decoupled, except to the 
extent that they interact via their effect on the worldsheet location 
itself whenever the codimension $q$ is non zero, or alternatively, if 
the codimension is zero, is zero, via their gravitational
coupling, which can in that case can be be easily incorporated (by 
requiring the background field $\Go_{\mu\nu}$ to satisfy Einstein's 
equations or some generalisation thereof) but which is not
so easy to deal with in a higher dimensional background (for which 
suitable methods of regularisation \cite{CBU03,BCM05} may be needed).

The other components of the tensor $\LL_{ab}$ in the separated
system will be given by
{\be \LL_{_{00}}=\frac{\partial \LL_{\rm_{S}}}{\partial \gg^{_{00}}}
=-\frac{\partial \LL_{\rm_{S}}}{\partial (\muv^2)}
\, ,\fe}
and by
{\be  \LL_{_{AB}}=\frac{\partial \LL_{\rm_{N}}}
{\partial \gg^{_{AB}}}=\frac{_1}{^2}(\LL_{\rm_{N}}\,\gamm_{_{AB}}-
 \PP_{_{AB}})\, ,\fe}
where $\gamma_{_{AB}}$ is the covariant inverse of the
contravariant base space metric $\gamm^{_{AB}}=\gg^{_{AB}}$ 
defined by (\ref{space}), and  $\PP_{_{AB}}$ will be
interpretable as the correspondingly index lowered version of the pressure
pressure tensor $\PP^{_{AB}}$ of the medium, from which its 
elasticity tensor will be obtainable in the usual way \cite{C73} as
{\be\EE_{\ \ \ _{CD}}^{\,_{AB}}=2\,\frac{\partial \PP^{_{AB}}}
{\partial \gg^{_{CD}}}- \PP^{_{AB}}\,\gamm_{_{CD}}\, .\fe}
This means that with respect to arbitrary worldsheet coordinates 
the pressure tensor $\PP_{ab}=\PP_{_{AB}}\vphi^{_A}_{\, ,a}
\vphi^{_B}_{\, ,b}$ of the elastic medium will have a contravariant 
version expressible as
{\be \PP^{ab}= \LL_{\!_{N}}\gamm^{ab}
-2\,\gamm^{ac}\gamm^{bd}\LL_{cd}\fe}
while the  ordinary elasticity tensor of the medium will be given by
{\be \EE^{abcd}=\LL_{\rm_{med}}(\gamm^{ab}\gamm^{bc}-2\gamm^{a(c}
\gamm^{d)b}) +\PP^{a(c}\gamm^{d)b}+\gamm^{a(c}\PP^{d)b}
-4\gamma^{ae}\gamma^{bf}\gamma^{cg}\gamma^{dh}\LL_{efgh}\, .\fe}

It follows from (\ref{stren}) that the total stress energy tensor 
will be given by the sum
{\be\TT^{ab}=\TT_{\!\rm_{S}}^{\, ab}+\TT_{\!\rm_{N}}^{\, ab}
\, ,\fe}
in which the scalar field contribution will be given by
{\be \TT_{\!\rm_{S}}^{\,ab}=\LL_{\rm_{S}}\,\gg^{ab}-2\LL_{\rm_{S}}^{\, ab}
\, ,\hskip 1 cm \LL_{\rm_{S}}^{\, ab}=- \LL_{\rm_{S}}^\prime
\muv^a \muv^b\, ,\hskip 1 cm\LL_{\rm_{S}}^\prime=
\frac{\partial \LL_{\rm_{S}}}
{\partial (\muv^2)}\, ,\fe}
which means that the scalar field will have a pressure $\PP_{\rm_S}$
and rest frame energy density $\rrho_{\rm_S}$ given by
{\be \PP_{\rm_S}=\LL_{\rm_S}\, ,\hskip 1 cm \rrho_{\rm_S}=
2\muv^2 \LL_{\rm_S}^\prime-\LL_{\rm_S}\, ,\label{sdens}\fe}
while the contribution from the ``normal'' part will be given by an 
expression of the standard form
{\be \TT_{\!\rm_{N}}=\rrho_{\rm_N}\, \uu^a \uu^b+\PP^{ab}\, ,\hskip 1 cm
\rrho_{\rm_N}=-\LL_{\rm_{N}}\, .\label{ndens}\fe}

\bigskip
{\bf 6. Separated characteristic equations.}
\medskip

Like the stress tensor, so also the  hyperelasticity tensor will be 
expressible in the separated case as a sum
{\be\CC^{abcd}=\CC_{\!\rm_{S}}^{\, abcd}+
\CC_{\!\rm_{N}}^{\, abcd}\, ,\fe}
in which, according to (\ref{hyp}), the scalar field contribution 
will be given by
{\be\CC_{\!\rm_{S}}^{\,abcd}=2 (\LL_{\!\rm_{S}}^{\,abcd}\!+
\!\LL_{\rm_{S}}^{\,a(c}\gg^{d)b}\!+\!
\gg^{a(c}\LL^{d)b})\!-\!(\LL_{\rm_{S}}^{\,ab}\gg^{cd}\!+\!\gg^{ab}
\LL_{\rm_{sca}}^{\,cd})\!+\!\frac{\LL_{\rm_{S}}}{2}(\gg^{ab}
\gg^{cd}\!-\!2\gg^{a(c}\gg^{d)b})\, ,\fe}
with
{\be   \LL_{\rm_{S}}^{abcd}\!=\LL_{\rm_{S}}^{\prime\prime}
\muv^a\muv^b\muv^c\muv^d\, ,\hskip 1 cm
\LL_{\rm_{S}}^{\prime\prime}=\frac{\partial\LL_{\rm_{S}}^\prime}
{\partial (\muv^2)}\, ,\fe}
while for the contribution from the ``normal'' elastic medium it can
be seen that we shall recover the Friedman Schutz \cite{FS75}  formula
{\be\CC_{\!\rm_{N}}^{\,abcd}=-\frac{_1}{^2} \EE^{abcd}\!+
\!\PP^{a(c}\uu^{d)}\uu^b\!+\!\uu^a\uu^{(c}\PP^{d)b}\!-\frac{_1}{^2}
\!(\PP^{ab}\uu^c\uu^d\!+\!\uu^a\uu^b\PP^{cd})\!+\!\frac{_1}{^2}
\rrho(\gg^{ab}\gg^{cd}\!-\!2\gg^{a(c}\gg^{d)b})\, ,\fe}

In the separated case the crossed element, with respect to the
preferred coordinate system,  of the characteristic
matrix (\ref{charm}) will automatically vanish 
{\be \Qq_{_{0A}}=0\, ,\fe}
and there will be a decoupling of the ``normal'' characteristic modes, 
for which the discontinuity amplitude $\ze^a$ is orthogonal to $\mu_a$, 
from the ``scalar'' characteristic mode for which $\ze^a$ is aligned 
with $\uu^a$. The latter is given by
{\be \ze^{_A}=0 \hskip 1 cm \Rightarrow \hskip 1 cm \Qq_{_{00}}=0
\, ,\fe}
so that one obtains a scalar characteristic equation in the form
{\be (2\LL_{\rm_S}^{\prime\prime}(\muv^a\lamda_a)^2-\LL_{\rm_S}^\prime
\lamda^2\, ,\hskip 1 cm \lamda^2=\gg^{ab})\lamda_a\lamda_b=0\, .\fe}
As the relative propagation speed $\vv_{\rm_S}$ will be definable
using a standard calibration of  the characteristic covector
$\lamda_a$ by
{\be \vv_{\rm_S}^{\,2}=(\muv^a\lamda_a)^2/\muv^2\, ,\hskip 1 cm
\lamda^2=1- \vv_{\rm_S}^{\,2}\, ,\fe}
it can be seen that the characteristic equation just gives the
condition
{\be \vv_{\rm_S}^{-2}=1+2\muv^2\LL_{\rm_S}^{\prime\prime}/
\LL_{\rm_S}^\prime\, .\label{scavel}\fe}
It can thus be seen that for a scalar model of the ``standard''
kind with a linear equation of state, meaning one with 
$\LL_{\rm_S}^{\prime\prime}=0$, the discontinuities will travel at 
the speed of light, $\vv_{\rm_S}^{\,2}=1$, while for other models
of  k-essence \cite{AMS00} or more general kinds
\cite{ECSAM02} it can be seen that the condition for causality, 
$\vv_{\rm_S}^{\,2}\leq 1$, and the reality condition for
local stability, $\vv_{\rm_S}^{\,2}\geq 0$, will both be
satisfied if and only if the scalar equation of state is such that
{\be 2\muv^2 \LL_{\rm_S}^{\prime\prime}/\LL_{\rm_S}^\prime\geq 0
\, .\fe}

As well as these simple ``scalar'' characteristic modes (not to 
mention the extrinsic characteristic modes that may exist, subject to 
(\ref{extrin}), if there is a higher dimensional background) there will
be different kinds of ``normal'' characteristic modes given by
{\be \ze^{_0}=0 \hskip 1 cm \Rightarrow \hskip 1 cm 
\Qq_{_{AC}}\ze^{_C}=0\, ,\label{norcha}\fe}
so that using the standard calibration
{\be \lamda_a=\nnu_a +\vv\uu_a\, , \hskip 1 cm
\nnu_a\uu^a=0\, , \hskip 1 cm \nnu^a\nnu_a=1\, ,\fe}
to define the relative propagation velocity $\vv$ and propagation
direction $\nnu^a$ (modulo a sign that may be chosen to make $\vv$ 
positive) the corresponding characteristic condition on the normal 
covector $\lamda_a$ will be the vanishing of the p-dimensional 
determinant of the matrix whose components can be seen from 
(\ref{charmat}) to be given by
{\be \Qq_{_{AC}}=2(\LL_{_{AC}}\lamda^2+2\LL_{_{ABCD}}\nnu^{_B}\nnu^{_D})
\, ,\hskip 1 cm \lamda^2=1-\vv^2\, .\fe}

For the simple fluid case -- as given by the restriction that
$\LL_{\rm_{N}}$ should depend only on the components of the base metric 
$\gamma_{_{AB}}$ only via its determinant $|\gamm|$ -- one will obtain
{\be \LL_{_{AB}}=-\LL_{\!\rm_{N}}^\wr\gamm_{_{AB}}\, ,\hskip 1 cm
\LL_{\!\rm_{N}}^\wr=\frac{\partial\LL_{\!\rm_{N}}}
{\partial({\rm ln}|\gamm|)}=|\gamm|\frac{\partial\LL_{\!\rm_{N}}}
{\partial|\gamm|}\, ,\fe}
so the pressure tensor will have the isotropic form, 
{\be \PP^{ab}=\PP_{\!\rm_{N}}\gamm^{ab}\, ,\fe}
in which the ordinary pressure scalar will be given by the well 
known (but in my previous review \cite{C83} miscopied) formula
{\be \PP_{\!\rm_{N}}=\LL_{\!\rm_{N}}+2\LL_{\!\rm_{N}}^\wr 
\, .\fe}
In terms of the corresponding bulk modulus, namely
{\be \Bbeta_{\!\rm_{N}}=-2 \PP_{\!\rm_{N}}^\wr=
-2|\gamm|\frac{\partial\PP_{\!\rm_{N}}}{\partial|\gamm|}
=-2\LL_{\!\rm_{N}}^\wr-4\LL_{\!\rm_{N}}^{\wr\wr}\, ,\fe}
 one will obtain
{\be \LL_{_{ABCD}}=-\frac{_1}{^4}\Bbeta\,\gamm_{_{AB}}\gamm_{_{CD}}
+\frac{_1}{^2}\LL_{\!\rm_{N}}^\wr\gamm_{_{AC}}\gamm_{_{BD}}+
\LL_{\!\rm_{N}}^\wr\gamm_{_{A[D}}\gamm_{_{B]C}}
\, ,\fe}
and the elasticity tensor will be given \cite{C83} by
{\be \EE^{abcd}=(\Bbeta_{\!\rm_{N}} -
\PP_{\!\rm_{N}})\gamm^{ab}\gamm^{cd}+2\PP_{\!\rm_{N}}
\gamm^{a(c}\gamm^{d)b}\, .\fe}

It can thus be seen that -- as in the more general case of an 
isotropic solid configuration \cite{BCCM05} -- the 
characteristic equation (\ref{norcha}) for the ``normal'' fluid will 
have solutions of two kinds, namely longitudinal modes with  
$\vv=\vv_{\rm L}$, and transverse modes with $\vv=\vv_{\rm S}$, of 
which the latter are non propagating in the fluid -- as opposed to 
solid --  case,
{\be \ze^{_A}\nnu_{_A}=0 \hskip 0.6 cm  \Rightarrow \hskip 0.6 cm 
\vv^2=\vv_{\rm S}^{\,2}\, ,\hskip 1.2 cm \vv_{\rm S}^{\,2}=0\, ,\fe}
while the former are just ordinary sound waves,
{\be \ze^{_A}=\nnu^{_A}\hskip 0.6 cm  \Rightarrow \hskip 0.6 cm
 \vv^2=\vv_{\rm L}^{\,2}\, ,\hskip 1 cm \vv_{\rm L}^{\,2}
=\Bbeta_{\rm_N}/(\rrho_{\rm_N}+\PP_{\rm_N})\, .\fe}

\bigskip
{\bf 7. Discussion: polytropic and poly-essential equations of state.}
\medskip

For both a ``normal'' fluid and a ``scalar'' constituent, an important 
role is played by their respective pressure to density ratios, namely
{\be \ww_{\rm N}=\PP_{\rm_N}/\rrho_{\rm_N}\, ,\hskip 1 cm
\ww_{\rm S}=\PP_{\rm_S}/\rrho_{\rm_S}\, ,\fe}
which can be seen to be given by
{\be \ww_{\rm N}=1+2\LL_{\!\rm_{N}}^\wr/\LL_{\!\rm_{N}} \, ,\hskip
1 cm \ww_{\rm S}^{-1}=2\muv^2\LL_{\!\rm_{S}}^\prime/\LL_{\!\rm_{S}} 
-1\, .\label{sound}\fe}
In recent years it has become common, in specialised cosmological 
(though not general astrophysical) literature, to use a rather loose 
terminology whereby the pressure to density ratio is referred to as the 
``equation of state'', an appelation that is jusifiable only if the 
ratio in question is actually constant. If -- as will often but not
always be a good approximation -- it can be supposed that this
ratio,  $\ww_{\rm N}$ or $\ww_{\rm S}$ as the case may be, is constant
then it will indeed characterise a corresponding equation of state.
In the ``normal'' fluid case, such an equation of state  will be of what 
is known (in the astrophysical literature) as polytropic type, with
polytropic index $\gamma_{\rm N}$, as given an  ansatz of the form
{\be \LL_{\rm_N}=-C |\gamm|^{-\gamma_{\rm N}/2}\, ,
\hskip 1 cm \gamma_{\rm N}= 1+\ww_{\rm N}\, ,\label{polytrope}\fe}
for some constant coefficient $C$. In the ``scalar'' case, the postulate 
that $\ww_{\rm S}$ should be constant can be seen to imply that the 
equation of state will be of what may be termed poly-essential type, as 
given by a power law ansatz, with index $\alpha_{\rm S}$, of the form
{\be \LL_{\rm_S}={\calK}\, \mu^{\alpha_{\rm S}}\, ,\hskip 1 cm
\alpha_{\rm S}=1+\ww_{\rm S}^{-1}\, ,\label{poless}\fe}
in which the coefficient ${\calK}$ is independent of the field gradient
magnitude $\muv$,  but is given as some function just of the
scalar field magnitude $\vphi^{_0}$. This means that the 
poly-esssential ansatz (\ref{poless})  characterises a special 
subcategory within the category of quintessential models called 
k-essential \cite{AMS00,G03,SV04}. This subcategory includes a model of 
the ``standard'' type, namely the trivial case of a free massless scalar 
field, in the limit when $\ww_{\rm S}=1$, which corresponds to 
$\alpha_{\rm S}=2$.

Attention in cosmology has centered during recent years on the
observational evidence to the effect that the universe is
accelerating, which suggests the need \cite{CMMS04} for a model 
with a rather strongly negative value, somewhere in the range
between -1/3 and -1,  for the mean pressure to density ratio $\ww$.
This poses a problem for a ``normal'' fluid model of the
kind specified for a given value of $\ww_{\rm N}$ by the equation
of state (\ref{polytrope}) for which it is well known that
the longitudinal modes (\ref{sound}) will have propagation velocity
given by
{\be \vv_{\rm L}^{\,2}=\ww_{\rm N}\, , \fe}
so that the conditions of reality (for local stability) and causality
imply the restrictions
{\be 0\leq \ww_{\rm N}\leq 1\, .\label{restn}\fe}

It has been rather unfairly suggested that, compared with such ordinary
fluid models, k-essential and other scalar models are advantaged by the 
absence of such a restriction, since they have squared characteristic 
velocity, $\vv_{\rm S}^{\,2}$ -- as given by equation (7) of the article 
\cite{AMS00} referred to -- that ``can be positive for any'' value
of the relevant ``equation of state'' ratio, $\ww_{\rm S}$. 

The reason why this is unfair is that if $\ww_{\rm S}$ is a bona fide
``equation of state'' parameter, meaning that it is actually constant,
then the corresponding equation of state, namely the poly-essential
ansatz (\ref{poless}) entails, according to (\ref{scavel}), that that 
the relevant propagation velocity will be given by
{\be \vv_{\rm S}^{\,2}=(\alpha_{\rm S}-1)^{-1}=\ww_{\rm S}\, ,\fe}
which means that the situation will just the same as in the
ordinary fluid case, in so much as the corresponding restriction
{\be 0\leq \ww_{\rm S}\leq 1\, .\label{restriction}\fe}
will have to be satisfied. 
 
The way suggested by Bucher and Spergel \cite{BS99} for getting over 
the restriction (\ref{restn}) is to a seek a mechanism providing elastic 
rigidity.  As an alternative, the way the advocates of k-essential 
(and other quintessential) cosmological models \cite{ECSAM02} propose 
to get around the restriction (\ref{restriction}) is of course to drop 
the postulate that $\ww_{\rm S}$ should be a genuine ``equation of 
state'' parameter, and instead use equations of state of more general 
kinds in which $\ww_{\rm S}$ is demoted from the status of a fixed 
parameter to that of a variable field. What is unfair is to give the 
impression that the possibity of doing that is a privilege 
distinguishing scalar field models from ``normal'' fluid models: in 
fact the use of more general kinds of ordinary fluid model -- in which 
the density to pressure ratio $\ww_{\rm N}$ is just a variable field 
that may be quite different from the squared sound speed -- is actually 
commonplace in many areas of astrophysics, and should not be 
prematurely ruled out of consideration in a cosmological context.

\bigskip
{\bf 8. Conclusions}
\medskip

The category of models presented in the preceeding work is
sufficient for a wide range of cosmological applications of the 
traditional 3+1  kind in which the codimension $q$ is taken to 
be zero, so that the spacetime metric can be taken to governed by 
the ordinary Einstein equations.
  
The treatment here has been set up in such a way as to be 
applicable also to scenarios in which our 4-dimensional spacetime 
manifold is considered to be a thin worldbrane in a higher -- 
meaning $4+q$ -- dimensional background with a given geometry. 
However for that kind of application it is difficult -- despite the
efforts of many workers in recent years -- to see how to allow 
realistically for the effect of gravity except in a Cowling type 
approximation in which the scale of the perturbations is supposed to 
be sufficiently small (compared with the relevant Jeans length) for 
their self interactions to be neglected. A much studied (albeit 
only marginally plausible) possibility -- that is beyond the scope of 
the present treatment but that does include allowance for strong 
gravitational coupling -- is that of the Randall-Sundrum $q=1$ type 
scenarios \cite{BDL00,STM00} (and their reflection symmetry violating 
generalisations \cite{BCMU01,KV02}) but the unsatisfactory features 
of such scenarios include the loss of effective predictability due to 
incoming gravitational waves from the bulk. 

In scenarios with codimension $q\geq 2$ it is even harder to see 
how to allow properly for gravitation. Attempts to account for the 
appearance of 4-dimensional gravity as a simulated effect 
\cite{CUBM01} due to acceleration with respect to a fixed bulk 
geometry tend to predict comportment of scalar - tensor type rather 
than the pure spin - 2 gravity that is actually observed. Progress has
however been achieved \cite{CBU03, BCM05} in the regularisation 
of the divergences that will result from true gravity in the bulk, 
and that can be allowed for (if not too strong) within a treatment 
of the kind presented here as an extra contribution to the effective 
stress energy tensor appearing in the extrinsic characteristic
equation (\ref{extrin}). The condition that the extrinsic propagation 
velocities should always be real and compatible with causality will 
evidently restrict the admissible values of the eigenvalues of the 
net stress-energy tensor in  (\ref{extrin}). In particular, if it 
is of the isotropic form
 {\be \TT^{\mu\nu}=\rho\left((1+\ww)\uu^\mu\uu^\nu +\ww
\,\gg^{\mu\nu}\right) \,\fe}
then the extrinsic propagation velocity $\vv_{\rm E}$ will be given by
{\be \vv_{\rm E}^{\,2}= -\ww \, .\fe}
It follows (unless the codimension $q$ is zero) that instead of 
being subject to a positivity condition of the familiar kind 
exemplified by (\ref{restn}) and (\ref{restriction}) the net 
pressure to density ratio $\ww$ will have to satisfy the negativity 
condition
{\be -1\leq\ww\leq 0\, .\fe}
Such pressure negativity is not incompatible with the observational 
evidence, but does exclude the possibility of describing the universe 
just in terms of scalar fields and fluids that are purely  of
respectively poly-essential and polytropic type.

\bigskip
\noindent
{\bf {Acknowledgements}}
\medskip

The author is grateful for discussions with  D. Steer about symplectic 
structures, with V. Mukhanov about non-standard inflation,  and with 
R. Battye and E. Chachoua about the elastic solid model of the universe.

\end{document}